\documentclass[12pt,preprint]{aastex}

\newcommand{\teff}{$T_{\!\mbox{\tiny\rm eff}}$}

\newcommand{\fglr}{\sc fglr\rm}
\newcommand{\wlr}{\sc wlr\rm}

\slugcomment{}

\shorttitle{SUPERGIANT VARIABILITY}
\shortauthors{Bresolin et al.}

\begin{document}



\title{On the photometric variability of blue supergiants in NGC~300 and its
impact on the Flux-weighted Gravity--Luminosity
Relationship$^1$}\footnotetext[1]{Based on observations
obtained with the 2.2m ESO/MPI telescope at the European Southern
Observatory, as part of proposal 163.N-0210.}

\author{Fabio Bresolin} \affil{Institute for Astronomy, 2680 Woodlawn
Drive, Honolulu, HI 96822; bresolin@ifa.hawaii.edu}

\author{Grzegorz Pietrzy\'nski}
\affil{Universidad de Concepci\'on, Departamento de Fisica, Casilla 160-C,
Concepci\'on, Chile and\\ Warsaw University Observatory, Al.
Ujazdowskie 4, 00-478, Warsaw, Poland; pietrzyn@hubble.cfm.udec.cl}

\author{Wolfgang Gieren} \affil{Universidad de Concepci\'on,
Departamento de Fisica, Casilla 160-C, Concepci\'on, Chile;
wgieren@coma.cfm.udec.cl}

\author{Rolf-Peter Kudritzki, Norbert Przybilla} \affil{Institute for Astronomy, 2680
Woodlawn Drive, Honolulu HI 96822; kud@ifa.hawaii.edu,
nob@ifa.hawaii.edu}

\and

\author{Pascal Fouqu\'{e}} \affil{Observatoire de Paris, LESIA, 5
Place Jules Janssen, F-92195 Meudon Cedex, France; pfouque@eso.org}

\begin{abstract}
We present a study of the photometric variability of spectroscopically
confirmed supergiants in NGC~300, comprising 28 epochs extending over
a period of five months. We find 15 clearly photometrically variable
blue supergiants in a sample of nearly 70 such stars, showing maximum
light amplitudes ranging from 0.08 to 0.23 magnitudes in the {\em V}\/
band, and one variable red supergiant. We show their light curves, and
determine semi-periods for two A2\,Ia stars. Assuming that the
observed changes correspond to similar variations in the bolometric
luminosity, we test for the influence of this variability on the
Flux-weighted Gravity--Luminosity Relationship and find a negligible
effect, showing that the calibration of this relationship, which has
the potential to measure extragalactic distances at the Cepheid
accuracy level, is not affected by the stellar photometric variability
in any significant way.

\end{abstract}



\section{Introduction}

The photometric and spectral variability has been established as a
characterizing feature of supergiant stars already a few decades ago
(\citealt{abt57}, \citealt{rosendhal70}, \citealt{maeder72},
\citealt{rosendhal73}, \citealt{sterken77}), affecting virtually all
luminous stars during their post-main sequence evolution. The light
variations appear to be cyclic, characterized by semi-regular, rather
than strictly periodic, patterns, obeying a
semi-period--luminosity--color relationship (\citealt{burki78},
\citealt{maeder80}) analogous to the relation valid for Cepheids, with
timescales ranging from a few days (early B supergiants) to hundreds
of days (red supergiants). The maximum light amplitudes, on the order
of 0.1 mag, increase with stellar luminosity, and reach in the hot
star domain a local maximum for early-B supergiants
(\citealt{appenzeller72}, \citealt{maeder80},
\citealt{vangenderen89}). The presence of photometric micro-variations
on scales $\Delta m<0.2$ mag typifies variable massive OBA
supergiants, or $\alpha$~Cyg variables following the General Catalogue
of Variable Stars (\citealt{kholopov85}; see also reviews by
\citealt{sterken89} and \citealt{vangenderen91,vangenderen01}).

The existence of a semi-period--luminosity--color relationship
suggests pulsational instabilities as the driving mechanism for the
observed variability, where stochastic processes determine the
quasi-regularity.  Several clues, including the large discrepancy
between the theoretical periods for radial pulsations in the
fundamental mode and the observed ones (\citealt{percy83},
\citealt{vangenderen85}), together with color variations smaller than
predicted (\citealt{vangenderen86}), indicate that non-radial
oscillations are at work in blue supergiants (\citealt{maeder86},
\citealt{baade92}), possibly appearing 
in connection with evolutionary stages past the red supergiant phase
(\citealt{lovey84},
\citealt{schaller90}). Alternative or additional mechanisms have been
proposed to explain the observed photometric and spectroscopic
variability, such as orbital motions in binary systems
(\citealt{harmanec87}). In the case of late-B and early-A supergiants
the variability of spectral line profiles originating in the
photosphere is consistent with non-radial pulsation modes, while
rotational modulation appears to be affecting the extended stellar
wind regions (\citealt{kaufer96,kaufer97}).

Regardless of the physical reasons for the observed variability, which
concerns both the photospheric layers (as deduced from the photometric
variations) and the extended winds of blue supergiants (from time
series spectroscopic investigations, e.g.~\citealt{kaufer96},
\citealt{rivinius97}, \citealt{stahl03}), it is interesting and
necessary to determine its relevance when considering the use of blue
supergiant stars as stellar candles (\citealt{bresolin03}). Recently
\citet{kudritzki03} have discussed the preliminary calibration of the
Flux-weighted Gravity--Luminosity Relationship (\fglr) and its
potential use as an extragalactic distance indicator. As originally
proposed, such a relation applies to late-B to early-A supergiants,
which belong to the visually brightest stars in galaxies, thus
allowing tests of its validity to be carried out in galaxies located
well beyond the Local Group. Both the \fglr\/ and the Wind
Momentum--Luminosity Relationship (\wlr, \citealt{kudritzki99}) rely
on the measurement of the apparent visual magnitude of blue
supergiants, which are then transformed to bolometric luminosities
from knowledge of the spectral type. An assessment of the
reliability of such methods should involve tests concerning their
sensitivity to time-dependent stellar properties, such as the
mass-loss rate (for the \wlr), and to the above mentioned photometric
micro-variations (\fglr\/ and \wlr).

`Normal' (non-LBV) blue supergiants have been rarely monitored for
variability beyond the Magellanic Clouds.  In this paper we report on
a medium-term (five months) {\em V}\/ monitoring program of blue
supergiants in the galaxy NGC~300. The main motivation for this work
is to demonstrate that the \fglr\/ is insensitive for all practical
purposes to the micro-variations normally observed in B and A
supergiants.

\section{Observational data}

The photometric measurements for the NGC~300 blue supergiants included
in this study have been extracted from images obtained at the ESO/MPI
2.2\,m telescope on La Silla, equipped with its Wide Field Imager
8\,K\,$\times$\,8\,K mosaic camera. These multi-epoch observations are
part of the {\em BVRI} Cepheid monitoring program described by
\citet{pietrzynski02}, where details on the calibration and the data
reduction can be found. The data discussed in the current paper span
the period 1999 July 31--2000 January 8, covered by observations on 28
different nights, mostly under photometric conditions, roughly grouped
in five blocks of data separated by temporal gaps up to 43 days
long. Only the {\em V}-band temporal sequence will be discussed here
({\em B}-band magnitudes have a somewhat less reliable calibration,
but still show variations similar to those observed in {\em V}).

Spectra of blue supergiants in NGC~300 have been obtained with the
FORS1 multi-object spectrograph at the ESO Very Large Telescope by
\citet{bresolin02}, and we restrict the following analysis to the
spectroscopically confirmed B- and A-type supergiants. For the
individual stars we adopt the nomenclature introduced in the spectral
catalog presented in that paper.

\section{Variability of blue supergiants: detection and light curves}

\begin{figure}
\plotone{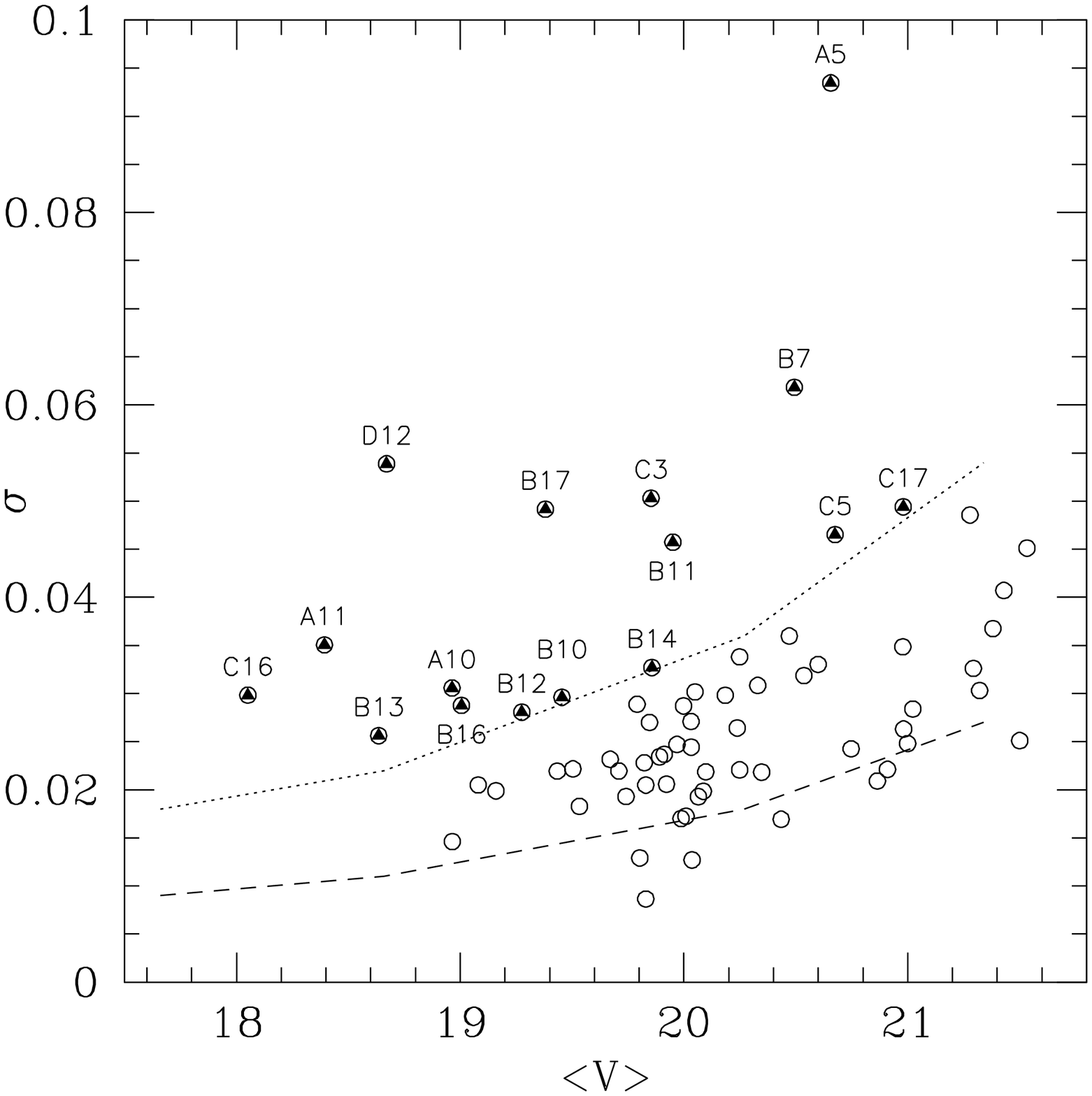}
\caption{Standard deviation from the mean magnitude as a
function of mean $V$ magnitude for the blue supergiant sample studied
spectroscopically in NGC~300. The dashed line indicates the standard
deviation measured for comparison stars, $\sigma_0$, while the dotted
line is plotted at $\sigma=2\,\sigma_0$. Stars above this limit are
shown with filled symbols, and are labeled according to the spectral
catalog of
\citet{bresolin02}.}
\label{sigma}
\end{figure}

The standard deviation from the mean magnitude and the maximum light
amplitude were taken as a measure of the stellar variability. We
selected stars having small dispersion in magnitude to serve as
comparison stars in order to define the observational uncertainty,
$\sigma_0$, as a function of magnitude. We then arbitrarily retained
for further analysis only those objects for which
$\sigma\ge2\,\sigma_0$ (Fig.~\ref{sigma}). A selection based on the
maximum light amplitude generates a similar sample of stars. This
procedure is justified since we are not aiming at a complete
compilation of variable supergiants in NGC~300, but instead at an
analysis of those spectroscopically confirmed supergiants displaying
the largest variability, since these will have a maximum effect on the
\fglr.

The 16 stars satisfying the selection criterion are labeled in
Fig.~\ref{sigma} with the identification number introduced by
\citet{bresolin02}. Their mean apparent {\em V} and absolute $M_V$
magnitudes, $B-V$ colors, standard deviations in the $V$ magnitude
$\sigma_{\!\mbox{\tiny\rm V}}$, maximum light amplitudes
A$_{\!\mbox{\tiny\rm V}}$ and spectral types are summarized in
Table~\ref{variables}. $M_V$'s have been determined assuming a
distance modulus $m-M=26.53$ (\citealt{freedman01}), and estimating
the reddening (and the extinction with $R_V=3.1$) based on the
difference between the observed $B-V$ and the expected color index for
the given spectral type.  The photometric measurements differ slightly
(mostly at the 0.01-0.02 mag level) from those in Table~2 of
\citet{bresolin02}, as a result of improved zero-points and
photometry. All stars in Table~\ref{variables} are blue supergiants
(early-B to mid-A), with the exception of A5, a red supergiant, which
shows the largest light variation among the detected variables,
apparently occuring over a period longer than our observing
window. The WN11 (or Ofpe/WN9) star B16 analyzed by
\citet{bresolin02b} also appears in Table~\ref{variables}.

\begin{figure}
\plottwo{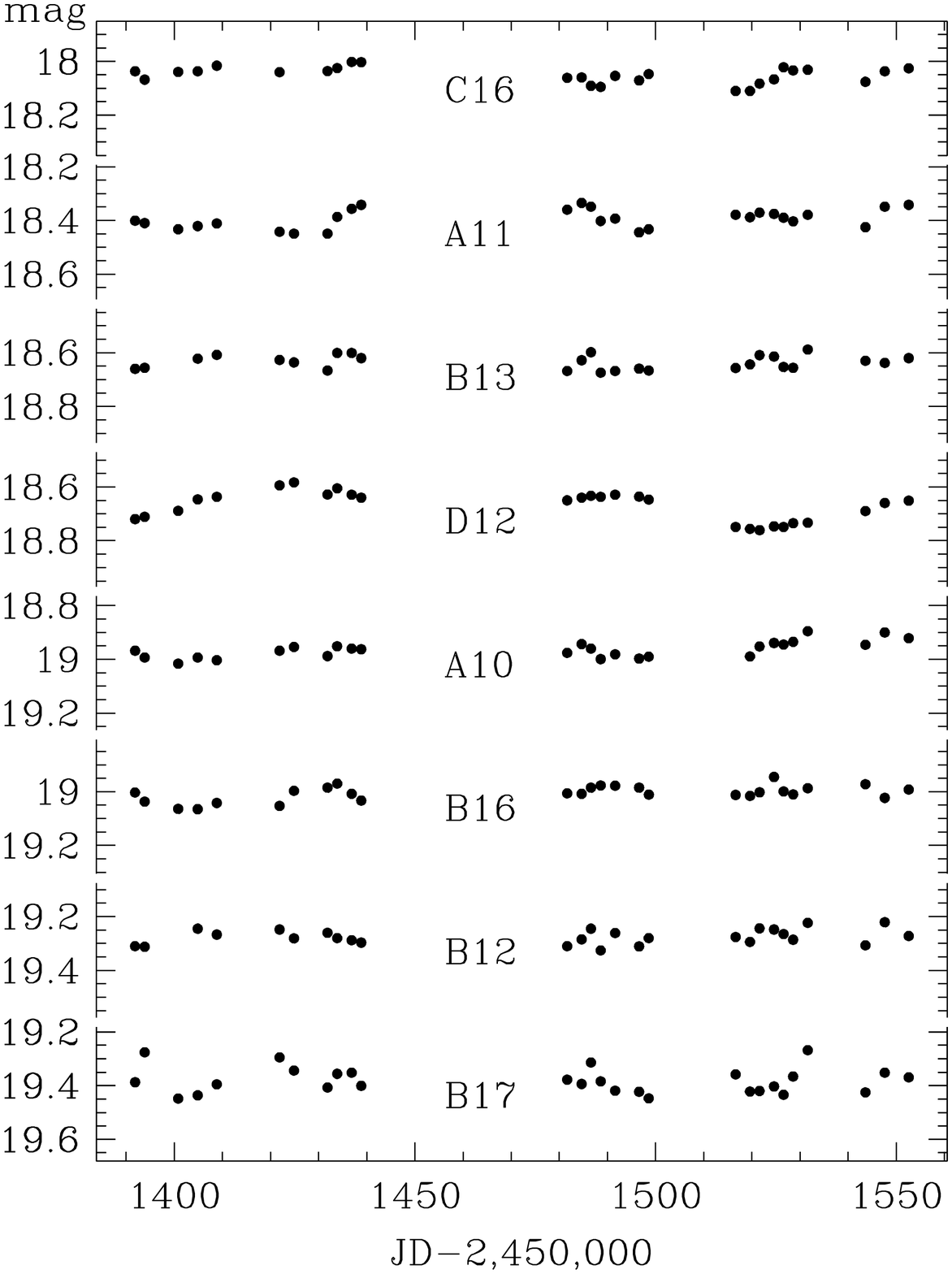}{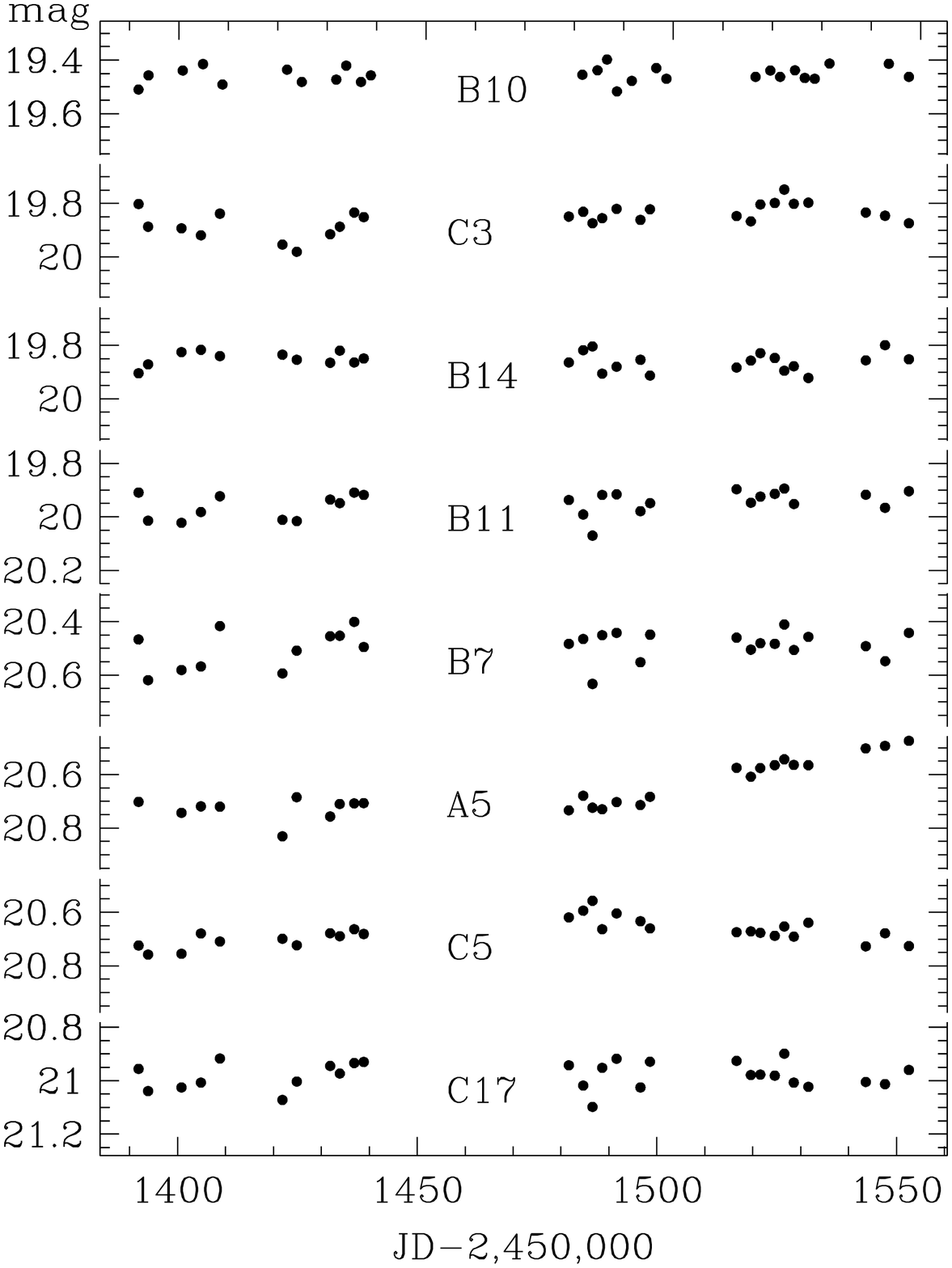}
\caption{$V$-band  light curves of the selected variable blue supergiants in
NGC~300, in order of increasing magnitude.}
\label{lightcurve}
\end{figure}

The {\em V}\/ light curves of the selected variable stars are displayed
in Fig.~\ref{lightcurve}. Despite the gaps in the temporal sequence,
photometric variations occuring on timescales of tens of days can be
easily detected in most cases, often superimposed on variations at
higher frequency.

\section{Period search}
The periodicity of the photometric sequences has been analyzed
utilizing the Phase Dispersion Minimization algorithm introduced by
\citet{stellingwerf78}, as implemented in the {\sc iraf} astronomical
data reduction package. Our goal was simply to check whether
semi-periods comparable in length to those of Galactic and Magellanic
Cloud supergiants could be found also in NGC~300, aware of the fact
that the irregular sampling of the data within our observing window is
far from being optimal for an accurate frequency analysis.

\begin{figure}
\plotone{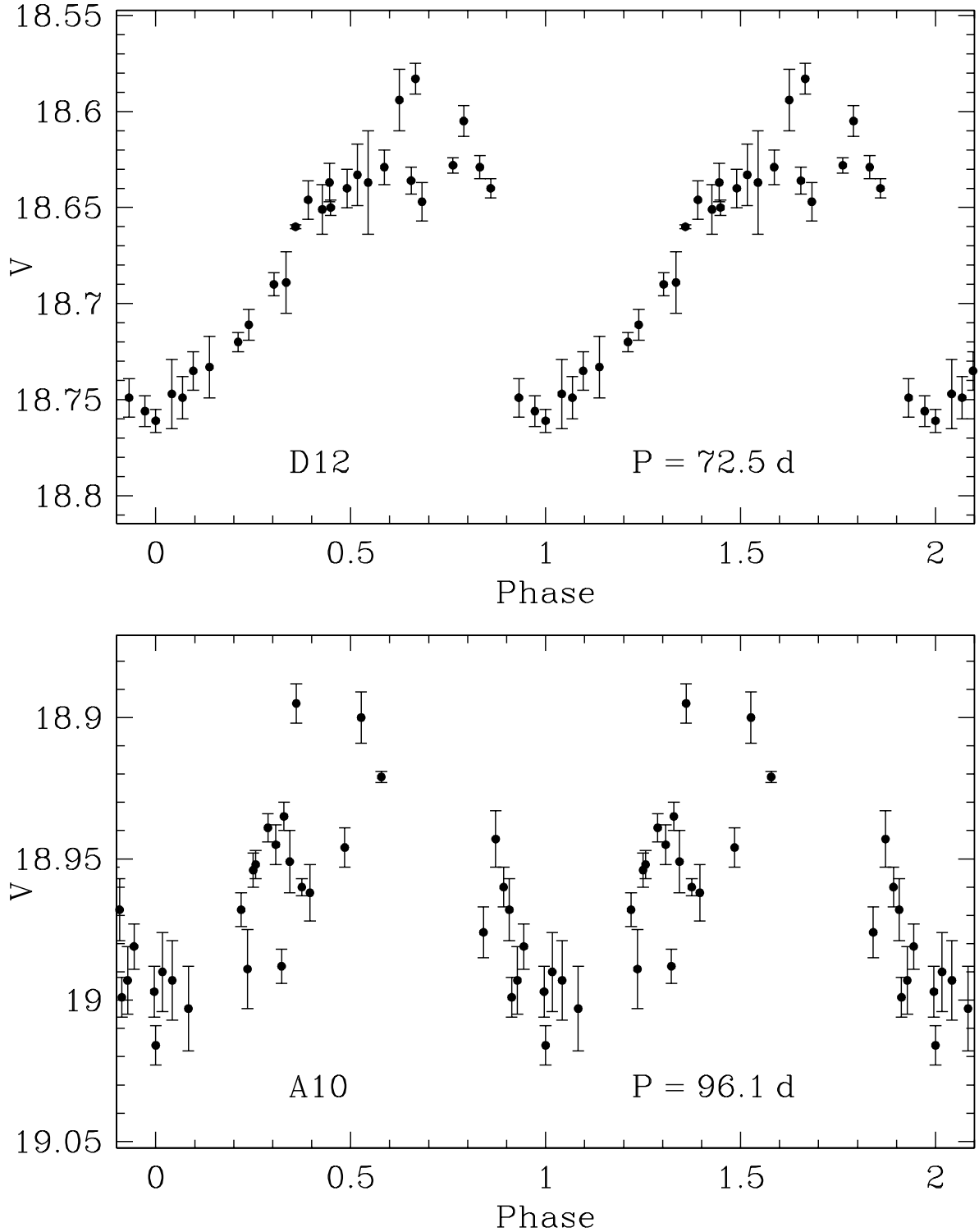}
\caption{$V$-band phase diagrams for the A2\,Ia stars D12 and A10,
showing periodic or near-periodic photometric variations. For clarity
the phase has been extended over two periods.}
\label{phase}
\end{figure}

Only for two stars a clear isolated minimum in the $\rm\Theta$
statistics (which is defined to vary between 0 and 1, see
\citealt{stellingwerf78}) is unambiguously identified, leading to a
reasonable determination of the semi-periods. Such is the case of D12
($\rm\Theta=0.06$, $\rm P=72$ days) and A10 ($\rm\Theta=0.40$, $\rm
P=96$ days), both of spectral type A2\,Ia, shown in the phase diagrams
of Fig.~\ref{phase} (the uncertainty in the quoted periods is on the
order of 10\%). For the remaining stars in the sample $\rm\Theta$ lies
between 0.4 and 0.75, and a unique semi-period determination was not
possible, with multiple phase dispersion minima occuring at periods
roughly between a few days and 30-40 days. The phased light curves for
these stars display a large amount of scatter, but this is typical of
the light variations of blue supergiants, as multiple pulsation modes
can be excited in blue supergiants (\citealt{lucy76},
\citealt{vangent86}, \citealt{vangenderen98}).  We note that the
semi-periods found for the two A2 supergiants D12 and A10, of absolute
magnitude $M_V=-8.3$ and $-7.8$, respectively, are roughly twice as
long as those listed for similar stars in the Milky Way by
\citet{burki78}. Fundamental periods calculated from the respective
stellar parameters are 42 days (D12) and 27 days (A10)
(\citealt{schaller90}).

\section{Impact on the Flux-weighted gravity--Luminosity Relationship}

The findings of the previous sections confirm that the typical
photometric micro-variability of blue supergiants, known from studies
in the Milky Way and the Magellanic Clouds, is also found in our
NGC~300 sample. The advantage of these data from the extragalactic
point of view is that they allow us to carry out a simple but
important test concerning the effect of the stellar variability on the
\fglr. Given the typical amplitude of the broad-band photometric
variations, $\sim\,$0.1 mag, it is reasonable not to expect a
significant effect, but a quantification of its size is desirable.

The \fglr\/, introduced by \citet{kudritzki03}, expresses the
proportionality between the bolometric magnitude of a blue supergiant
and its flux-weighted gravity during its evolution across the
Hertzprung-Russell diagram at roughly constant luminosity:

\begin{equation}
M_{\!\mbox{\tiny\rm bol}} = 3.71 \, \log(g/T_{\!\mbox{\tiny \rm
eff,4}}^4)\, - 13.49 
\end{equation}

\noindent
where $T_{\!\mbox{\tiny \rm eff,4}}=T_{\!\mbox{\tiny \rm
eff}}/10^4\,K$, with slope and zero-point fixed by a preliminary
calibration, based on the stellar parameters determined for stars in
different galaxies: Milky Way, Magellanic Clouds, NGC~6822, M31, M33,
NGC~300 and NGC~3621.

It is not clear whether during the semi-periodic light variations the
bolometric magnitude remains rigorously constant. If that were the
case, then the \fglr\/ would be unaffected by the blue supergiant
variability. On the other hand, one could imagine that the detected
light variations are symptomatic of small changes in the total
luminosity (as in the pulsating massive star models by
\citealt{dorfi00}).  In this case, and lacking additional constraints,
we can test the corresponding effect on the \fglr\/ by determining the
relationship at different epochs, letting $M_{\!\mbox{\tiny\rm bol}}$
vary by an amount equal to the broad-band variability.  The
flux-weighted gravity $g/T_{\!\mbox{\tiny\rm eff,4}}^4$ is kept
constant, since spectra for the derivation of gravity and \teff\/ are
available for one single epoch only, and therefore we have no way of
estimating their possible variations.  Only spectral time series would
allow us to verify if and how the stellar parameters change during the
cyclic changes.

Under these assumptions, we have calculated slope and zero-point of
the \fglr\/ at each available epoch, using our photometric temporal
sequence for the NGC~300 blue supergiants considered by Kudritzki et
al.~(2003), i.e.~14 stars of spectral type between B8 and A2.  The
stellar parameters for these stars, as calculated by Kudritzki et
al.~(2003), are summarized in Table~\ref{fglrstars}, and the whole set
of $V$ magnitudes at 28 different epochs is given in Table~3 (the
uncertainty of the individual measurements is on the order of
0.01--0.02 mag).  {\em The zero-point and rms of the linear fit are
found to be affected below the 10\% level, when compared with the
regression shown for NGC~300 stars alone ($rms\simeq\,$0.2 mag) by
Kudritzki et al.~(2003), with variations in the slope only up to
4\%}. It is clear that the supergiant variability introduces a
negligible effect on the resulting \fglr, which will remain negligible
even when an improved calibration of the relation, including its
possible dependence on the stellar metallicity, will become available
in the near future.

In conclusion, the \fglr\/ appears to provide an extragalactic
distance indicator which is robust against supergiant photometric
variability, provided that the number of stars considered per galaxy
is sufficiently large ($n\sim\,$10). As for the future work, besides
improving the calibration in terms of statistics and metallicities,
additional efforts are required to establish whether spectral
variations are important for the derivation of the stellar parameters.

\acknowledgments WG gratefully acknowledges financial support for this
work from the Chilean Center for Astrophysics FONDAP 15010003. Support
from the Polish KBN grant 2P03D02123 is acknowledged.




\clearpage

\begin{deluxetable}{ccccccc}

\tabletypesize{\scriptsize}
\tablecolumns{7}
\tablewidth{0cm}
\tablecaption{Selected variables\label{variables}}

\tablehead{
\colhead{Supergiant}     &
\colhead{$<\!V\!>$}             &
\colhead{$<\!M_V\!>$\phantom{$^{\rm a}$}}             &
\colhead{\phantom{$-$}$<\!B-V\!>$}           &
\colhead{$\sigma_{\!\mbox{\tiny\rm V}}$}        &
\colhead{A$_{\!\mbox{\tiny\rm V}}$}             &
\colhead{Spectral}      \\
\colhead{ID}    &
\colhead{}      &
\colhead{}      &
\colhead{}      &
\colhead{}      &
\colhead{}      &
\colhead{type}}
\startdata
A5\dotfill   &   20.66  &  $-$6.02$^{\rm a}$  & \phantom{$-$}2.04 & 0.093 & 0.357 & Late\hspace*{\fill}  \\
A10\dotfill  &   18.96  &  $-$7.72\phantom{$^{\rm a}$}  & \phantom{$-$}0.07 & 0.031 & 0.121 & A2\hspace*{\fill}  \\ 
A11\dotfill  &   18.39  &  $-$8.51\phantom{$^{\rm a}$}  & \phantom{$-$}0.10 & 0.035 & 0.114 & B8\hspace*{\fill}   \\
B7\dotfill   &   20.49  &  $-$6.29\phantom{$^{\rm a}$}  & $-$0.08 & 0.062 & 0.232 & B2\hspace*{\fill}  \\ 
B10\dotfill  &   19.45  &  $-$7.20\phantom{$^{\rm a}$}  & \phantom{$-$}0.06 & 0.030 & 0.119 & A2--A3\hspace*{\fill}  \\ 
B11\dotfill  &   19.95  &  $-$6.58\phantom{$^{\rm a}$}  & \phantom{$-$}0.04 & 0.046 & 0.176 & A5\hspace*{\fill}   \\
B12\dotfill  &   19.28  &  $-$7.56\phantom{$^{\rm a}$}  & $-$0.11 & 0.028 & 0.105 & B0.5\hspace*{\fill}  \\ 
B13\dotfill  &   18.63  &  $-$7.90\phantom{$^{\rm a}$}  & $-$0.08 & 0.026 & 0.086 & B3\hspace*{\fill}   \\
B14\dotfill  &   19.86  &  $-$6.73\phantom{$^{\rm a}$}  & \phantom{$-$}0.04   & 0.033 & 0.123 & A0\hspace*{\fill}   \\
B16\dotfill  &   19.01  &  $-$7.72\phantom{$^{\rm a}$}  & $-$0.07 & 0.029 & 0.120 & WN11\hspace*{\fill}   \\
B17\dotfill  &   19.38  &  $-$7.58\phantom{$^{\rm a}$}  & $-$0.07 & 0.049 & 0.180 & B0.5\hspace*{\fill}  \\ 
C3\dotfill   &   19.85  &  $-$6.93\phantom{$^{\rm a}$} & \phantom{$-$}0.06   & 0.050 & 0.233 & B8--B9\hspace*{\fill}   \\
C5\dotfill   &   20.67  &  $-$6.20\phantom{$^{\rm a}$}  & $-$0.05  & 0.046 & 0.201 & B2\hspace*{\fill}   \\
C16\dotfill  &   18.05  &  $-$8.70\phantom{$^{\rm a}$}  & \phantom{$-$}0.07   & 0.030 & 0.108 & B9\hspace*{\fill}  \\
C17\dotfill  &   20.98  &  $-$5.70$^{\rm a}$  & \phantom{$-$}0.14  &  0.049 & 0.199 & Composite\hspace*{\fill}  \\
D12\dotfill  &   18.67  &  $-$8.32\phantom{$^{\rm a}$}  & \phantom{$-$}0.18  & 0.054 & 0.178 & A2 \hspace*{\fill}  
\enddata
\tablenotetext{a}{assuming $E(B-V)=0.05$}

\end{deluxetable}

\begin{deluxetable}{cccccc}

\tabletypesize{\scriptsize}
\tablecolumns{6}
\tablewidth{0cm}
\tablecaption{{\sc fglr} supergiants in NGC 300 -- Stellar parameters\label{fglrstars}}

\tablehead{
\colhead{Supergiant}     &
\colhead{Spectral}             &
\colhead{\teff}             &
\colhead{$\log g$}           &
\colhead{$E(B-V)$}        &
\colhead{Bolometric}             \\
\colhead{ID}    &
\colhead{type}      &
\colhead{($K$)}      &
\colhead{}      &
\colhead{}      &
\colhead{correction}}      
\startdata
A8\dotfill   &   B9-A0  &  10,000  &  1.60  & 0.02  &  $-0.42$  \\
A10\dotfill   &   A2     &  9,000   &  1.30  & 0.05  &  $-0.22$  \\
A11\dotfill   &   B8     & 12,000   &  1.42  & 0.12  &  $-0.91$  \\
A13\dotfill   &   B8     & 12,000   &  1.95  & 0.02  &  $-0.81$  \\
A18\dotfill   &   B8     & 12,000   &  1.90  & 0.05  &  $-0.83$  \\
B10\dotfill   &   A2-A3  &  8,800   &  1.37  & 0.04  &  $-0.14$  \\
C6\dotfill   &   A0     &  9,500   &  1.60  & 0.07  &  $-0.28$  \\
C7\dotfill   &   A0     &  9,500   &  1.68  & 0.03  &  $-0.31$  \\
C8\dotfill   &   B8     & 12,000   &  1.90  & 0.03  &  $-0.83$  \\
C12\dotfill   &   B9-A0 &  9,800   &  1.80  & 0.06  &  $-0.40$  \\
C16\dotfill   &   B9     & 10,500   &  1.30  & 0.07  &  $-0.56$  \\
D12\dotfill   &   A2     &  9,000   &  1.05  & 0.15  &  $-0.27$  \\
D13\dotfill   &   A0     &  9,500   &  1.35  & 0.03  &  $-0.31$  \\
D17\dotfill   &   B9-A0  & 10,000   &  1.60  & 0.05  &  $-0.39$  
\enddata

\end{deluxetable}

\begin{deluxetable}{ccccccccccccccc}

\tabletypesize{\scriptsize}
\tablecolumns{15}
\tablewidth{0cm}
\rotate
\tablecaption{{\sc fglr} supergiants in NGC 300 -- $V$-magnitude temporal sequence}

\tablehead{
\colhead{Supergiant}     &
\multicolumn{14}{c}{Julian Date $-$ 2,450,000}	\\
\colhead{ID}    &
\colhead{1391.86}	&
\colhead{1393.86}	&
\colhead{1400.81}	&
\colhead{1404.88}	&
\colhead{1408.85}	&
\colhead{1421.87}	&
\colhead{1424.85}	&
\colhead{1431.83}	&
\colhead{1433.83}	&
\colhead{1436.84}	&
\colhead{1438.84}	&
\colhead{1481.61}	&
\colhead{1484.64}	&
\colhead{1486.59}	\\
\colhead{}	&
\colhead{1488.57}	&
\colhead{1491.59}	&
\colhead{1496.57}	&
\colhead{1498.59}	&
\colhead{1516.62}	&
\colhead{1519.59}	&
\colhead{1521.60}	&
\colhead{1524.60}	&
\colhead{1526.57}	&
\colhead{1528.58}	&
\colhead{1531.60}	&
\colhead{1543.57}	&
\colhead{1547.61}	&
\colhead{1552.58}}
\startdata
A8\dotfill & 19.45 & 19.48 & 19.46 & 19.44 & 19.44 & 19.47 & 19.43 & 19.45 & 19.43 & 19.41 & 19.42 & 19.46 & 19.40 & 19.41 \\
           & 19.45 & 19.43 & 19.45 & 19.43 & 19.43 & 19.44 & 19.43 & 19.44 & 19.44 & 19.43 & 19.39 & 19.44 & 19.39 & 19.41 \\
A10\dotfill & 18.97 & 18.99 & 19.02 & 18.99 & 19.00 & 18.97 & 18.95 & 18.99 & 18.95 & 18.96 & 18.96 & 18.98 & 18.94 & 18.96 \\ 
           & 19.00 & 18.98 & 18.99 & 18.99 & \nodata & 18.99 & 18.95 & 18.94 & 18.94 & 18.93 & 18.89 & 18.95 & 18.90 & 18.92 \\
A11\dotfill & 18.40 & 18.41 & 18.43 & 18.42 & 18.41 & 18.44 & 18.45 & 18.45 & 18.39 & 18.36 & 18.34 & 18.36 & 18.33 & 18.35 \\
            & 18.40 & 18.39 & 18.44 & 18.43 & 18.38 & 18.39 & 18.37 & 18.38 & 18.39 & 18.40 & 18.38 & 18.42 & 18.35 & 18.34 \\
A13\dotfill & 19.84 & 19.83 & 19.83 & 19.83 & 19.83 & 19.82 & 19.80 & 19.83 & 19.81 & 19.79 & 19.82 & 19.84 & 19.81 & \nodata \\
            & 19.85 & 19.86 & 19.86 & \nodata & 19.83 & 19.82 & 19.81 & 19.82 & 19.83 & 19.83 & 19.76 & 19.84 & 19.77 & 19.82 \\
A18\dotfill & 20.01 & 20.00 & 20.01 & 19.99 & 19.98 & 19.96 & 19.98 & 20.06 & 20.01 & 20.01 & 20.01 & 20.00 & 19.95 & \nodata \\
            & 20.05 & 20.03 & 20.01 & 20.02 & 20.00 & 19.99 & 19.98 & 19.99 & 20.02 & 19.98 & 19.96 & 19.98 & 19.93 & 20.01 \\
B10\dotfill & 19.51 & 19.46 & 19.44 & 19.41 & 19.49 & 19.44 & 19.48 & 19.47 & 19.42 & 19.48 & 19.46 & 19.45 & 19.44 & 19.40 \\
            & 19.52 & 19.48 & 19.43 & 19.47 & 19.46 & 19.44 & 19.46 & 19.44 & 19.47 & 19.47 & 19.41 & 19.41 & 19.46 & \nodata \\
C6\dotfill &  19.90 & 19.95 & 19.94 & 19.94 & 19.88 & 19.90 & 19.92 & 19.91 & 19.90 & 19.88 & 19.89 & 19.92 & 19.92 & 19.94 \\
            & 19.93 & 19.88 & 19.95 & 19.91 & 19.89 & 19.91 & 19.90 & 19.93 & 19.87 & 19.91 & 19.88 & 19.92 & 19.93 & 19.92 \\
C7\dotfill &  20.34 & 20.38 & 20.37 & 20.34 & 20.35 & 20.37 & 20.34 & 20.35 & 20.33 & 20.32 & 20.32 & 20.36 & 20.35 & 20.38 \\
            & 20.38 & 20.33 & 20.39 & 20.35 & 20.33 & 20.35 & 20.33 & 20.35 & 20.32 & 20.36 & 20.31 & 20.34 & 20.34 & 20.33 \\
C8\dotfill &  20.01 & 20.09 & 20.07 & 20.08 & 20.03 & 20.09 & 20.05 & 20.04 & 20.05 & 20.01 & 20.01 & 20.03 & 20.02 & 20.11 \\
            & 20.06 & 20.01 & 20.10 & 20.04 & 20.03 & \nodata & 20.03 & 20.06 & 20.02 & 20.09 & 20.07 & 20.04 & 20.05 & 20.03 \\
C12\dotfill & 20.16 & 20.19 & 20.16 & 20.14 & 20.16 & 20.16 & 20.17 & 20.16 & 20.15 & 20.14 & 20.16 & 20.19 & 20.18 & 20.20 \\ 
            & 20.23 & 20.18 & 20.22 & 20.21 & 20.20 & 20.21 & 20.17 & 20.21 & 20.16 & 20.22 & 20.16 & 20.23 & 20.22 & 20.23 \\
C16\dotfill & 18.04 & 18.07 & 18.04 & 18.04 & 18.02 & 18.04 & \nodata & 18.04 & 18.02 & 18.00 & 18.00 & 18.06 & 18.06 & 18.09 \\
            & 18.09 & 18.05 & 18.07 & 18.05 & 18.11 & 18.11 & 18.08 & 18.07 & 18.02 & 18.03 & 18.03 & 18.08 & 18.04 & 18.03 \\
D12\dotfill & 18.72 & 18.71 & 18.69 & 18.65 & 18.64 & 18.59 & 18.58 & 18.63 & 18.60 & 18.63 & 18.64 & 18.65 & 18.64 & 18.63 \\
            & 18.64 & 18.63 & 18.64 & 18.65 & 18.75 & 18.76 & 18.76 & 18.75 & 18.75 & 18.73 & 18.73 & 18.69 & 18.66 & 18.65 \\
D13\dotfill & 18.98 & 18.97 & 18.99 & 18.94 & 18.96 & 18.98 & 18.96 & 18.99 & 18.97 & 18.97 & 18.96 & 18.97 & 18.96 & 18.96 \\
           & 18.96 & 18.98 & 18.97 & 18.98 & 18.95 & 18.94 & 18.95 & 18.96 & 18.98 & 18.97 & 18.94 & 18.95 & 18.95 & 18.96 \\
D17\dotfill & 19.50 & 19.51 & 19.52 & 19.53 & 19.58 & 19.53 & 19.52 & 19.55 & 19.53 & 19.53 & 19.53 & 19.54 & 19.54 & 19.53 \\
            & 19.57 & 19.57 & 19.54 & 19.55 & 19.51 & 19.52 & 19.53 & 19.53 & 19.52 & 19.51 & 19.51 & 19.53 & 19.53 & 19.55 \\
\enddata
\end{deluxetable}

\end{document}